\documentclass[twocolumn]{aastex63}

\usepackage{amsmath}
\usepackage{color}
\usepackage{url}
\usepackage{ulem}
\usepackage{multirow}
\usepackage{amsmath}
\usepackage{wasysym}

\usepackage{rotating, graphicx}

\def\apjl{ApJ}%
\def\apjs{ApJS}%
\def\ao{Appl.~Opt.}%
\def\aap{A\&A}%
\newcommand\C   {\textsc{Clumpy}}

\shorttitle{BASS XXXVII: radiation-regulated SMBH growth}
\shortauthors{Ricci et al.}

\begin{document}

\title{BASS XXXVII: The role of radiative feedback in the growth and obscuration properties of nearby supermassive black holes}

\author{C. Ricci}
\affiliation{N\'ucleo de Astronom\'ia de la Facultad de Ingenier\'ia, Universidad Diego Portales, Av. Ej\'ercito Libertador 441, Santiago, Chile} 
\affiliation{Kavli Institute for Astronomy and Astrophysics, Peking University, Beijing 100871, China}
\author{T. T. Ananna}
\affiliation{Department of Physics and Astronomy, Dartmouth College, 6127 Wilder Laboratory, Hanover, NH 03755, USA}
\author{M. J. Temple}
\affiliation{N\'ucleo de Astronom\'ia de la Facultad de Ingenier\'ia, Universidad Diego Portales, Av. Ej\'ercito Libertador 441, Santiago, Chile} 
\author{C. M. Urry}
\affiliation{Yale Center for Astronomy \& Astrophysics, Physics Department, PO Box 208120, New Haven, CT 06520-8120, USA}
\author{M. J. Koss}
\affiliation{Eureka Scientific, 2452 Delmer Street Suite 100, Oakland, CA 94602-3017, USA}
\affiliation{Space Science Institute, 4750 Walnut Street, Suite 205, Boulder, CO 80301, USA}
\author{B. Trakhtenbrot}
\affiliation{School of Physics and Astronomy, Tel Aviv University, Tel Aviv 69978, Israel}
\author{Y. Ueda}
\affiliation{Department of Astronomy, Kyoto University, Kyoto 606-8502, Japan}
\author{D. Stern}
\affiliation{Jet Propulsion Laboratory, California Institute of Technology, 4800 Oak Grove Drive, MS 169-224, Pasadena, CA 91109, USA}
\author{F. E. Bauer}
\affiliation{Instituto de Astrof\'isica, Facultad de F\'isica, Pontificia Universidad Cat\'olica de Chile, Campus San Joaquin, Av. Vicu\~na Mackenna 4860, Macul Santiago, Chile, 7820436}
\affiliation{Centro de Astroingenier\'ia, Facultad de F\'isica, Pontificia Universidad Cat\'olica de Chile, Campus San Joaquin, Av. Vicu\~na Mackenna 4860, Macul Santiago, Chile, 7820436}
\affiliation{Millennium Institute of Astrophysics, Nuncio Monse\~nor S\'otero Sanz 100, Of 104, Providencia, Santiago, Chile}
\author{E. Treister}
\affiliation{Instituto de Astrof\'isica, Facultad de F\'isica, Pontificia Universidad Cat\'olica de Chile, Campus San Joaquin, Av. Vicu\~na Mackenna 4860, Macul Santiago, Chile, 7820436}
\author{G. C. Privon}
\affiliation{National Radio Astronomy Observatory, 520 Edgemont Road, Charlottesville, VA 22903, USA}
\affiliation{Department of Astronomy, University of Florida, P.O. Box 112055, Gainesville, FL 32611, USA}
\author{K. Oh}
\affiliation{Korea Astronomy \& Space Science institute, 776, Daedeokdae-ro, Yuseong-gu, Daejeon 34055, Republic of Korea}
\author{S. Paltani}
\affiliation{Department of Astronomy, University of Geneva, ch. d'\'Ecogia 16, CH-1290 Versoix}
\author{M. Stalevski}
\affiliation{Astronomical Observatory, Volgina 7, 11060 Belgrade, Serbia}
\affiliation{Sterrenkundig Observatorium, Universiteit Ghent, Krijgslaan 281 S9, B-9000 Ghent, Belgium}
\author{L. C. Ho}
\affiliation{Kavli Institute for Astronomy and Astrophysics, Peking University, Beijing 100871, China}
\affiliation{Department of Astronomy, School of Physics, Peking University, Beijing 100871, China}
\author{A. C. Fabian}
\affiliation{Institute of Astronomy, Madingley Road, Cambridge CB3 0HA, UK}
\author{R. Mushotzky}
\affiliation{Department of Astronomy, University of Maryland, College Park, MD 20742, USA}
\affiliation{Joint Space-Science Institute, University of Maryland, College Park, MD 20742, USA}
\author{C. S. Chang}
\affiliation{Joint ALMA Observatory, Avenida Alonso de Cordova 3107, Vitacura 7630355, Santiago, Chile}
\author{F. Ricci}
\affiliation{Dipartimento di Fisica e Astronomia, Universita di Bologna, via Gobetti 93/2, I-40129 Bologna, Italy}
\affiliation{INAF Osservatorio Astronomico di Bologna, via Gobetti 93/3, I-40129 Bologna, Italy}
\author{D. Kakkad}
\affiliation{Space Telescope Science Institute, 3700 San Martin Drive, Baltimore, MD 21218, USA}
\author{L. Sartori}
\affiliation{Institute for Astronomy, Department of Physics, ETH Zurich,Wolfgang-Pauli-Strasse 27, CH-8093 Zurich, Switzerland}
\author{R. Baer}
\affiliation{Institute for Astronomy, Department of Physics, ETH Zurich,Wolfgang-Pauli-Strasse 27, CH-8093 Zurich, Switzerland}
\author{T. Caglar}
\affiliation{Leiden Observatory, PO Box 9513, 2300 RA, Leiden, the Netherlands}
\author{M. Powell}
\affiliation{Institute of Particle Astrophysics and Cosmology, Stanford University, 452 Lomita Mall, Stanford, CA 94305, USA}
\author{F. Harrison}
\affiliation{Cahill Center for Astronomy and Astrophysics, California Institute of Technology, Pasadena, CA 91125, USA}

\begin{abstract}

We study the relation between obscuration and supermassive black hole (SMBH) growth using a large sample of hard X-ray selected Active Galactic Nuclei (AGN). We find a strong decrease in the fraction of obscured sources above the Eddington limit for dusty gas ($\log \lambda_{\rm Edd}\gtrsim -2$) confirming earlier results, and consistent with the radiation-regulated unification model. This also explains the difference in the Eddington ratio distribution functions (ERDFs) of type\,1 and type\,2 AGN obtained by a recent study. The break in the ERDF of nearby AGN is at $\log \lambda_{\rm Edd}^{*}=-1.34\pm0.07$. This corresponds to the $\lambda_{\rm Edd}$ where AGN transition from having most of their sky covered by obscuring material to being mostly devoid of absorbing material. A similar trend is observed for the luminosity function, which implies that most of the SMBH growth in the local Universe happens when the AGN is covered by a large reservoir of gas and dust. These results could be explained with a radiation-regulated growth model, in which AGN move in the $N_{\rm H}$--$\lambda_{\rm Edd}$ plane during their life cycle. The growth episode starts with the AGN mostly unobscured and accreting at low $\lambda_{\rm Edd}$. As the SMBH is further fueled, $\lambda_{\rm Edd}$, $N_{\rm H}$ and covering factor increase, leading AGN to be preferentially observed as obscured. Once $\lambda_{\rm Edd}$ reaches the Eddington limit for dusty gas, the covering factor and $N_{\rm H}$ rapidly decrease, leading the AGN to be typically observed as unobscured. As the remaining fuel is depleted, the SMBH goes back into a quiescent phase.

\end{abstract}


\setcounter{footnote}{0}

\section{Introduction}

Supermassive black holes (SMBHs, with masses $M_{\rm BH}\geq 10^{6}M_{\odot}$) are found at the center of most massive galaxies (e.g., \citealp{Kormendy:1995fk}), and are thought to gain most of their mass through the accretion of matter from their circumnuclear environment (e.g., \citealp{Soltan:1982vc,Yu:2002ut,Shankar:2004cu}). During the rapid accretion phase, SMBHs can emit a large amount of radiation across the entire electromagnetic spectrum (e.g., \citealp{Elvis:1994wl}), outshining their host galaxies, and are observed as Active Galactic Nuclei (AGN).  The discovery of correlations between SMBH mass and several properties of their host galaxies, such as the luminosity and mass of the bulge (e.g., \citealp{Marconi:2003ov,Haring:2004xy}) and the velocity dispersion (e.g., \citealp{Ferrarese:2000uq,Gebhardt:2000kx,Kormendy:2013fk}), has suggested that AGN could play an important role in the evolution of galaxies. This is usually associated to a feedback process, in which the energy and radiation produced by the AGN interact with the interstellar medium (ISM) of their host galaxies (e.g., \citealp{Fabian:2012mk}), directly affecting the star formation process. Both semi-analytic models of galaxy formation (e.g., \citealp{Croton:2006xo,Bower:2006tm}) and hydrodynamical simulations (e.g., \citealp{Sijacki:2007xh,Schaye:2015nb}) have demonstrated the importance of AGN feedback, showing that such a mechanism is necessary to regulate star formation and to explain the high-mass end of the galaxy mass function. This feedback process could be associated to either radiative feedback (for luminous AGN; e.g., \citealp{Fabian:2012mk}) or to kinetic feedback (for low-luminosity AGN; e.g., \citealp{Weinberger:2017te}).

About 70$\%$ of the luminous\footnote{with a 14--150\,keV luminosity $\log (L_{14-150}/\rm erg\,s^{-1})\gtrsim 42.5$} AGN in the local Universe are found to be obscured by weakly-ionized or neutral gas [$\log (N_{\rm H}/\rm cm^{-2})\geq 22$; \citealp{Ricci:2015fk}], which implies that gas and dust typically cover a similar fraction of the sky as seen from the nucleus, assuming simple orientation-based unification (e.g., \citealp{Antonucci:1993vt,Urry:1995ga}). This obscuring material is thought to be located on circumnuclear scales, and distributed anisotropically around the AGN, leading to the classification of AGN into type\,2 (obscured) and type\,1 (unobscured) sources (e.g., \citealp{Antonucci:1993vt,Netzer:2015cx,Ramos-Almeida:2017ar,Hickox:2018jh}). Optical, UV and soft X-ray ($<10$\,keV) radiation can be strongly suppressed by line-of-sight obscuration, which leads to a strong bias against detecting heavily obscured sources in these energy bands. In the hard X-rays ($\geq 10$\,keV), obscuration is less important due to the lower photo-electric cross section of the obscuring material, which enables recovery of most of the X-ray flux up to $\log (N_{\rm H}/\rm cm^{-2})\simeq 23.5$ (e.g., \citealp{Ricci:2015fk}). Therefore, hard X-ray surveys, such as those carried out by {\it INTEGRAL} (e.g., \citealp{Paltani:2008bx,Beckmann:2009si,Krivonos:2022xr}), {\it Swift}/BAT (e.g., \citealp{Markwardt:2005cz,Tueller:2008eh,Tueller:2010po,Cusumano:2010bo,Baumgartner:2013uq,Oh:2018ie}) and {\it NuSTAR} (e.g., \citealp{Alexander:2013ac,Mullaney:2015gj,Civano:2015cd,Harrison:2016sj,Del-Moro:2017ly,Zappacosta:2018hi,Masini:2018bj}), are very well suited to detect and characterize obscured AGN, particularly at low redshift. The all-sky {\it Swift}/BAT survey, in particular, has detected $\sim$1,100\,AGN in the 14--195\,keV band \citep{Oh:2018ie}. X-ray follow-up of BAT-detected sources showed that a significant fraction ($\sim 20-30\%$) of local AGN are obscured by Compton-thick material [CT, $\log (N_{\rm H}/\rm cm^{-2})\geq 24$; e.g., \citealp{Burlon:2011dk,Ricci:2015fk,Akylas:2016uj,Marchesi:2018hw,Torres-Alba:2021kl,Tanimoto:2022id}].

The nuclear obscuring material can be significantly affected by the strong radiation emitted by the AGN (e.g., \citealp{Fabian:2006lq}). This feedback process was originally supported by a decrease of the fraction of obscured sources ($f_{\rm obs}$) with increasing luminosity. This was first discovered forty years ago by \citeauthor{Lawrence:1982bt} (\citeyear{Lawrence:1982bt}; see also \citealp{Lawrence:1991ah}), and then confirmed by numerous studies carried out in the optical (e.g., \citealp{Simpson:2005ac,Oh:2015tb}) and X-rays (e.g., \citealp{Ueda:2003qf,Ueda:2014ix,Steffen:2003df,La-Franca:2005ec,Barger:2005la,Treister:2006se,Hasinger:2008fj,Winter:2009ti,Brusa:2010wo}). Several works focussing on the IR regime also found evidence of a decrease of the covering factor of the obscurer with increasing luminosity (e.g., \citealp{Maiolino:2007ii,Treister:2008ff,Sazonov:2012gt,Lusso:2013pt,Stalevski:2016kl,Mateos:2017il,Lanz:2019jp,Ichikawa:2019zz,Toba:2021ys}). Several of these early results may however be affected by inconsistent bolometric corrections \citep{Netzer:2016jd}, and by the fact that they did not take into account the effects of anisotropy and radiation transfer \citep{Stalevski:2016kl}. When this is taken into account, the decrease of the covering factor with the AGN luminosity is reduced or disappears altogether \citep{Stalevski:2016kl,Netzer:2016jd}. As shown by \citet{Burlon:2011dk} in a study of {\it Swift}/BAT AGN, the relation between the fraction of sources with $\log (N_{\rm H}/\rm cm^{-2})=22-24$ and the luminosity is tightly connected to the different X-ray luminosity functions of obscured and unobscured AGN (see also \citealp{Della-Ceca:2008mh,Buchner:2015ve,Ananna:2019rz}). 

Studying a large number of {\it Swift}/BAT AGN, \citet{Ricci:2017rn} demonstrated that the main parameter driving the fraction of obscured sources is the Eddington ratio ($\lambda_{\rm Edd}$). This was done by showing the existence of a steep decrease of $f_{\rm obs}$ at $\log \lambda_{\rm Edd}\gtrsim -2$, which corresponds to the expected Eddington limit for dusty gas with $\log (N_{\rm H}/\rm cm^{-2})\simeq 22$ (e.g., \citealp{Fabian:2006lq,Fabian:2008hc,Fabian:2009ez,Ishibashi:2018ti}, see also \citealp{Honig:2007st} and \citealp{Kawakatu:2020nh}). \citet{Ricci:2017rn} also showed that, when controlling for $\lambda_{\rm Edd}$, the relation between $f_{\rm obs}$ and the AGN luminosity disappears. These results suggested that radiative feedback plays a dominant role in shaping the close environments of SMBHs, and led to the formulation of the radiation-regulated unification model \citep{Ricci:2017rn}, according to which the likelihood of a source to be observed as obscured is higher at low Eddington ratio ($\lambda_{\rm Edd}\lesssim -1.5$; Fig.\,4 of \citealp{Ricci:2017rn}). At higher Eddington ratios ($\lambda_{\rm Edd}\gtrsim -1.5$) the effect of radiation pressure clears the immediate vicinity of the AGN, possibly giving rise to the polar emission that has been observed in a large fraction of AGN in the mid-IR (e.g., \citealp{Tristram:2007kq,Honig:2013np,Honig:2017wm,Lopez-Gonzaga:2016cj,Asmus:2019wz,Alonso-Herrero:2021yl}). The fact that $\lambda_{\rm Edd}$ is the dominant parameter also points towards most of the obscuring material being located within the sphere of influence of the SMBH (typically $\lesssim 60$\,pc for the sample of \citealp{Ricci:2017rn} ), in agreement with recent studies carried out with ALMA (e.g., \citealp{Garcia-Burillo:2021ix}).

In this work we study the dependence of the fraction of AGN with a given $N_{\rm H}$ on the Eddington ratio, the relation between the AGN Eddington ratio distribution function and the covering factor of the circumnuclear material, and explain these relations with a radiation-regulated model for the growth of SMBHs. We use the second BASS data release (DR2; \citealp{Koss:2022nt}) to build a sample that contains $\sim 2$\,\,times more {\it Swift}/BAT AGN with black hole mass available with respect to the previous study of \citet{Ricci:2017rn}. The BASS DR2 sample is significantly more complete than the DR1 one, with 100\% of measured redshifts and $\sim 98\%$ of black hole masses for unbeamed AGN outside the Galactic plane. This allows us to study, for the first time, the relation between obscuration and $\lambda_{\rm Edd}$ in different ranges of $N_{\rm H}$. In a companion paper (Ananna et al. 2022b) we will present a complementary analysis of the sample, focussing on the timescales of the different stages of AGN growth. Throughout the paper we adopt standard cosmological parameters ($H_{0}=70\rm\,km\,s^{-1}\,Mpc^{-1}$, $\Omega_{\mathrm{m}}=0.3$, $\Omega_{\Lambda}=0.7$). All fractions are calculated following the Bayesian approach outlined in  \citet{Cameron:2011yw}, and the uncertainties quoted represent the 16th and 84th quantiles of a binomial distribution.

\begin{figure}
\centering
 %% 1st image
 %% 2nd image
\includegraphics[width=0.48\textwidth]{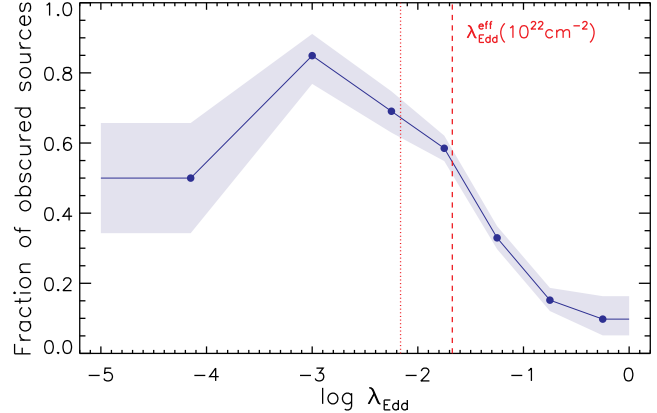}
% %% caption
% \begin{minipage}[t]{1\textwidth}
  \caption{Fraction of obscured Compton-thin [$22 \leq \log (N_{\rm H}/\rm cm^{-2})< 24$] sources versus Eddington ratio for AGN from our hard X-ray selected sample with $-4.8 \leq \log \lambda_{\rm Edd}< 0$. For each $\lambda_{\rm Edd}$ bin the fractions were normalized to unity in the $20 \leq \log (N_{\rm H}/\rm cm^{-2})< 24$ interval. The red lines show the expected Eddington limit for dusty gas with $\log (N_{\rm H}/\rm cm^{-2})\simeq 22$ from \citeauthor{Fabian:2006lq,Fabian:2008hc,Fabian:2009ez} (\citeyear{Fabian:2006lq,Fabian:2008hc,Fabian:2009ez}; dashed line) and \citeauthor{Ishibashi:2018ti} (\citeyear{Ishibashi:2018ti}; dotted line). The fractions are calculated following \citet{Cameron:2011yw}, and the uncertainties quoted represent the 16th and 84th quantiles of a binomial distribution.}
\label{fig:FobsVsEdd}
% \end{minipage}
\end{figure}

\begin{figure*}
\centering
 %% 1st image
 %% 2nd image
\includegraphics[width=0.95\textwidth]{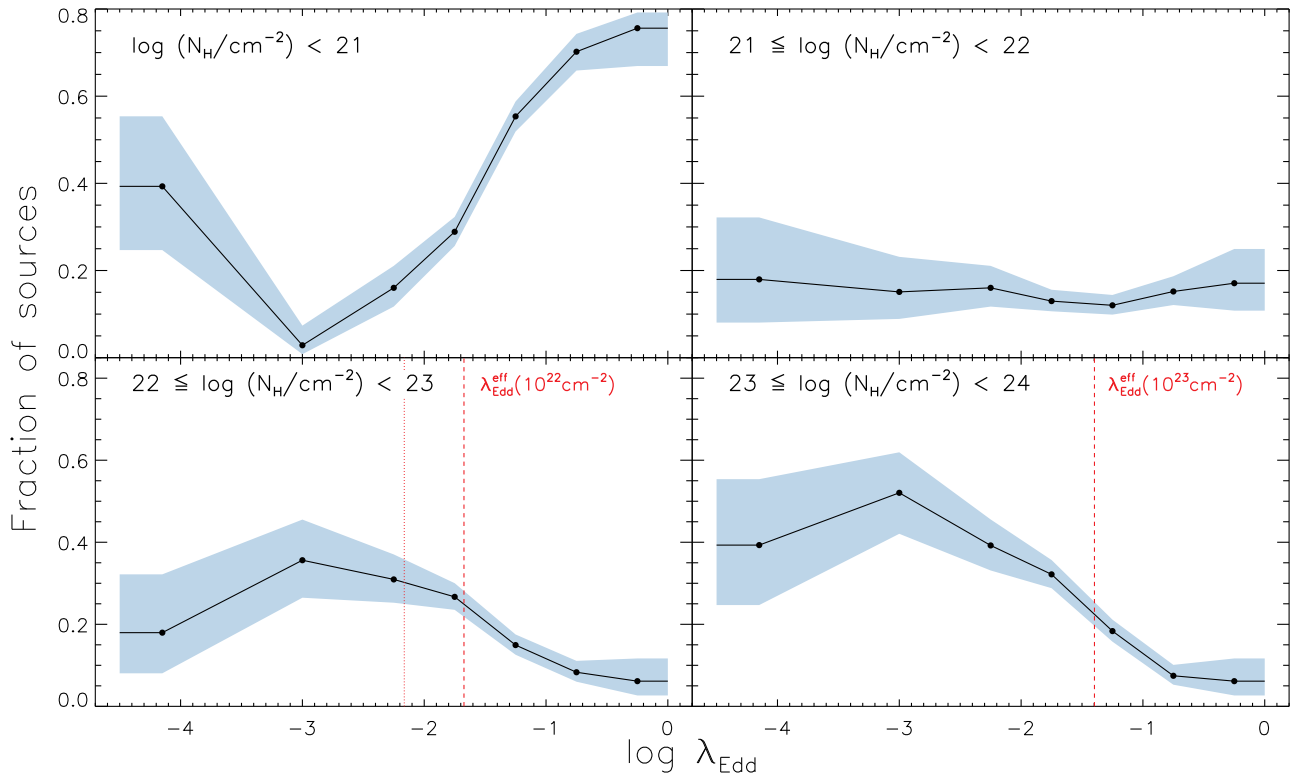}
% %% caption
% \begin{minipage}[t]{1\textwidth}
  \caption{ Fraction of sources with $N_{\rm H}$ in a given range versus Eddington ratio for the objects in our hard X-ray selected sample with $-4.5 \leq \log \lambda_{\rm Edd}< 0$. For each $\lambda_{\rm Edd}$ bin the fractions were normalized to unity in the $20 \leq \log (N_{\rm H}/\rm cm^{-2})< 24$ interval. The red lines show the expected Eddington limit for dusty gas with $\log (N_{\rm H}/\rm cm^{-2})\simeq 22$ (bottom left panel; dashed line from \citealp{Fabian:2006lq,Fabian:2008hc,Fabian:2009ez}, dotted line from \citealp{Ishibashi:2018ti}) and $\log (N_{\rm H}/\rm cm^{-2})\simeq 23$ (bottom right panel; \citealp{Ishibashi:2018ti,Venanzi:2020gx}). The latter value of the effective Eddington limit includes the contribution from IR radiation trapping. The fractions are calculated following \citet{Cameron:2011yw}, and the uncertainties quoted represent the 16th and 84th quantiles of a binomial distribution.}
\label{fig:FobsVsEdd_NHbins}
% \end{minipage}
\end{figure*}

\section{Sample}
The Burst Alert Telescope (BAT; \citealp{Barthelmy:2005uq}) on board the {\it Neil Gehrels Swift Observatory} \citep{Gehrels:2004dq} has been carrying out an all-sky survey in the 14--195\,keV band since its launch in November 2004. This has led to the detection of  more than 1,500 sources (e.g., \citealp{Barthelmy:2005uq,Oh:2018ie}), including over a thousand AGN. The BAT AGN Spectroscopic Survey (BASS\footnote{www.bass-survey.com}) has been gathering a large number of optical spectroscopy and ancillary multi-wavelength data for BAT-selected AGN. This includes data in the radio \citep{Baek:2019pp,Smith:2020ng}, mm \citep{Koss:2021zw,Kawamuro:2022mo}, infrared \citep{Ichikawa:2017zp,Ichikawa:2019zz,Lamperti:2017uz,den-Brok:2022oy,Ricci:2022eo}, optical \citep{Koss:2017wu} and X-rays \citep{Ricci:2017pm}. This has led to a number of follow-up studies comparing the X-ray continuum and optical properties of AGN with their accretion rates \citep{Trakhtenbrot:2017xm,Oh:2017di,Ricci:2018du,Rojas:2020or,Kakkad:2022xc}. The first BASS data release \citep{Koss:2017wu} reported black hole masses for 473 AGN, and X-ray properties for all 838 AGN from the flux-limited {\it Swift}/BAT 70-month sample \citep{Ricci:2017pm}. The second BASS data release \citep{Koss:2022nt} reported black hole masses for 780 AGN, and increased the total number of AGN from the {\it Swift}/BAT 70-month catalogue to 858.

In BASS black hole masses were obtained using single-epoch broad Balmer line measurements for unobscured AGN and velocity dispersions for obscured AGN. Typical systematic uncertainties on $M_{\rm BH}$ are $\sim 0.3-0.5$\,dex. In this work we use intrinsic X-ray fluxes and column densities from \citet{Ricci:2017pm}, and black hole masses from BASS DR2. Sources which were found to be unobscured in the X-ray band were assigned $\log (N_{\rm H}/\rm cm^{-2})= 20$ (i.e. an upper limit). To calculate the luminosities we used the updated distances and redshifts reported in \citep{Koss:2022qi}, which are based on emission line redshifts and redshift-independent distance measurements. Eddington ratios were calculated from the intrinsic X-ray luminosities as in \citet{Ricci:2017rn}. Similarly to what was done in \citet{Ricci:2017rn}, we excluded blazars from our sample \citep{Paliya:2019zz} as well as obscured objects for which the black hole mass was estimated using broad optical emission lines, since $M_{\rm BH}$ is typically underestimated for those objects \citep{Mejia-Restrepo:2022mt}. Our final sample consists of 681 AGN, spanning a large range in 14--150\,keV luminosities ($10^{41}-10^{45.5}\rm\,erg\,s^{-1}$), and with redshifts typically $z<0.15$ (with most of the objects being located within a few hundred Mpc).

\section{The relation between obscuration and Eddington ratio}\label{sec:fobsvsEdd}

\subsection{Obscured fraction vs Eddington ratio}
Figure\,\ref{fig:FobsVsEdd} shows the relation between the fraction of obscured Compton-thin [$\log (N_{\rm H}/\rm cm^{-2})= 22-24$] sources ($f_{\rm obs}$) and the Eddington ratio for our sample. We considered here only sources with $\log (N_{\rm H}/\rm cm^{-2})\leq 24$, to maximize the completeness of our selection (see Fig.\,1 of \citealp{Ricci:2015fk}). The observed trend reflects what was previously found by \citet{Ricci:2017rn}, with $f_{\rm obs}$ decreasing sharply at $\lambda_{\rm Edd}\gtrsim 10^{-2}$, a value consistent with the expected Eddington limit for dusty gas (red dashed lines; \citealp{Fabian:2006lq,Fabian:2008hc,Fabian:2009ez,Ishibashi:2018ti}). Consistent results have been recently obtained using different approaches and samples (see e.g., \citealp{She:2018vf} for a study of nearby low-luminosity AGN). From careful modelling of the broad-band X-ray spectra of nearby AGN using recently-developed torus models, \citet{Zhao:2020ql} and \citet{Ogawa:2021um} found a decrease of the covering factor of the obscuring material at $\lambda_{\rm Edd}\simeq 10^{-2}$. A decrease of the covering factor with increasing $\lambda_{\rm Edd}$ was also found in the IR (e.g., \citealp{Ezhikode:2017xq,Zhuang:2018cz}). This is in agreement with what would be expected by the radiation-regulated unification model \citep{Ricci:2017rn}, according to which the probability of observing a source as an obscured is a function of the inclination angle as well as the Eddington ratio.

Most of the sources in our sample are found to accrete at $\lambda_{\rm Edd}<1$, with only ten sources accreting at higher Eddington ratios. Although we have only a small number of sources at $\lambda_{\rm Edd}\geq 1$, it is interesting to notice that  three of them (LEDA\,97012, ESO\,383$-$18  and IRAS\,04210+0400) are obscured, which corresponds to $f_{\rm obs}=32^{+14}_{-12}\%$, a value significantly higher than what we found at $\lambda_{\rm Edd}\simeq 10^{-0.25}$ (see Fig.\,\ref{fig:FobsVsEdd}). Among these objects, LEDA\,97012 is in a merging system (e.g., \citealp{Koss:2012qf}), ESO\,383$-$18 shows peculiar absorption properties, possibly associated with partial covering \citep{Ricci:2010lu}, while IRAS\,04210+0400 shows a relatively high star formation rate (e.g., \citealp{Ichikawa:2017zp}). Larger studies of nearby AGN accreting at high Eddington ratios are needed to confirm an increase in $f_{\rm obs}$ at $\lambda_{\rm Edd}\gtrsim 1$.

\subsection{$f_{\rm obs}(N_{\rm H})$ vs $\lambda_{\rm Edd}$}

Considering the large number of AGN with careful measurements of black hole masses provided by BASS DR2 \citep{Koss:2022nt,Koss:2022uw}, we can now explore the trend between the fraction of sources in relatively narrow ranges of $N_{\rm H}$ and the Eddington ratio. Fig.\,\ref{fig:FobsVsEdd_NHbins} shows that the fraction of sources which show little to no absorption [$\log (N_{\rm H}/\rm cm^{-2})<21$] increases with $\lambda_{\rm Edd}$, and that such an increase is particularly steep at $\lambda_{\rm Edd}\gtrsim 10^{-2}$ (top left panel), in agreement with the idea that radiation pressure is able to clean up the obscuring material very rapidly, leaving most of the sky around the AGN with little or no absorbing material \citep{Ricci:2017rn}. A similar trend is obtained when selecting only sources with $20 < \log (N_{\rm H}/\rm cm^{-2})<21$. Interestingly, such a trend is not observed for sources with $\log (N_{\rm H}/\rm cm^{-2})= 21-22$ (top right panel), which exhibit a constant fraction of $\simeq 20\%$ across four orders of magnitude in $\lambda_{\rm Edd}$. This could be due to the fact that most of the obscuring material in this range of column densities is associated with gas from the host galaxy (e.g., \citealp{Buchner:2017jx,Malizia:2020ik}), and therefore is not affected by radiation pressure. In the $\log (N_{\rm H}/\rm cm^{-2})= 22-23$ range (bottom left panel) we observe a trend that is in good agreement with what is expected by considering the effect of radiation pressure, with a rapid decline at $\lambda_{\rm Edd}\simeq 10^{-2}$. Interestingly, the fraction of sources with $\log (N_{\rm H}/\rm cm^{-2})= 23-24$ (bottom right panel) shows a very similar trend to that of sources with $\log (N_{\rm H}/\rm cm^{-2})= 22-23$, with a rapid drop in the fraction of sources with Eddington ratios above a few percent. The Eddington limit for dusty gas is however expected to increase with the column density of the material, and it should be $\lambda_{\rm Edd}\simeq 10^{-1.15}$ for $\log (N_{\rm H}/\rm cm^{-2})\simeq 23$ (e.g., \citealp{Fabian:2006lq}). Recent theoretical studies (e.g., \citealp{Ishibashi:2018ti,Venanzi:2020gx}) have shown that infrared radiation trapping could play an important role in the obscuring material, and would allow AGN to expel dense [$\log (N_{\rm H}/\rm cm^{-2})\geq 23$] gas at relatively low Eddington ratios ($\lambda_{\rm Edd}\simeq 0.04$), which could explain our new observational results (dashed red vertical line in the bottom right panel of Fig.\,\ref{fig:FobsVsEdd_NHbins}). We explored the same trends of Fig.\,\ref{fig:FobsVsEdd_NHbins} by dividing our sample into different bins of intrinsic 14--150\,keV luminosity and black hole mass in Appendix\,\ref{sec:appendix1} (see Fig.\,\ref{fig:FobsVsEdd_NHbins_lumMBH}), and found that these parameters do not appear to have any significant effect on the relation between the fraction of sources with a given $N_{\rm H}$ and $\lambda_{\rm Edd}$.

\begin{figure}
 \centering
 \includegraphics[width=0.48\textwidth]{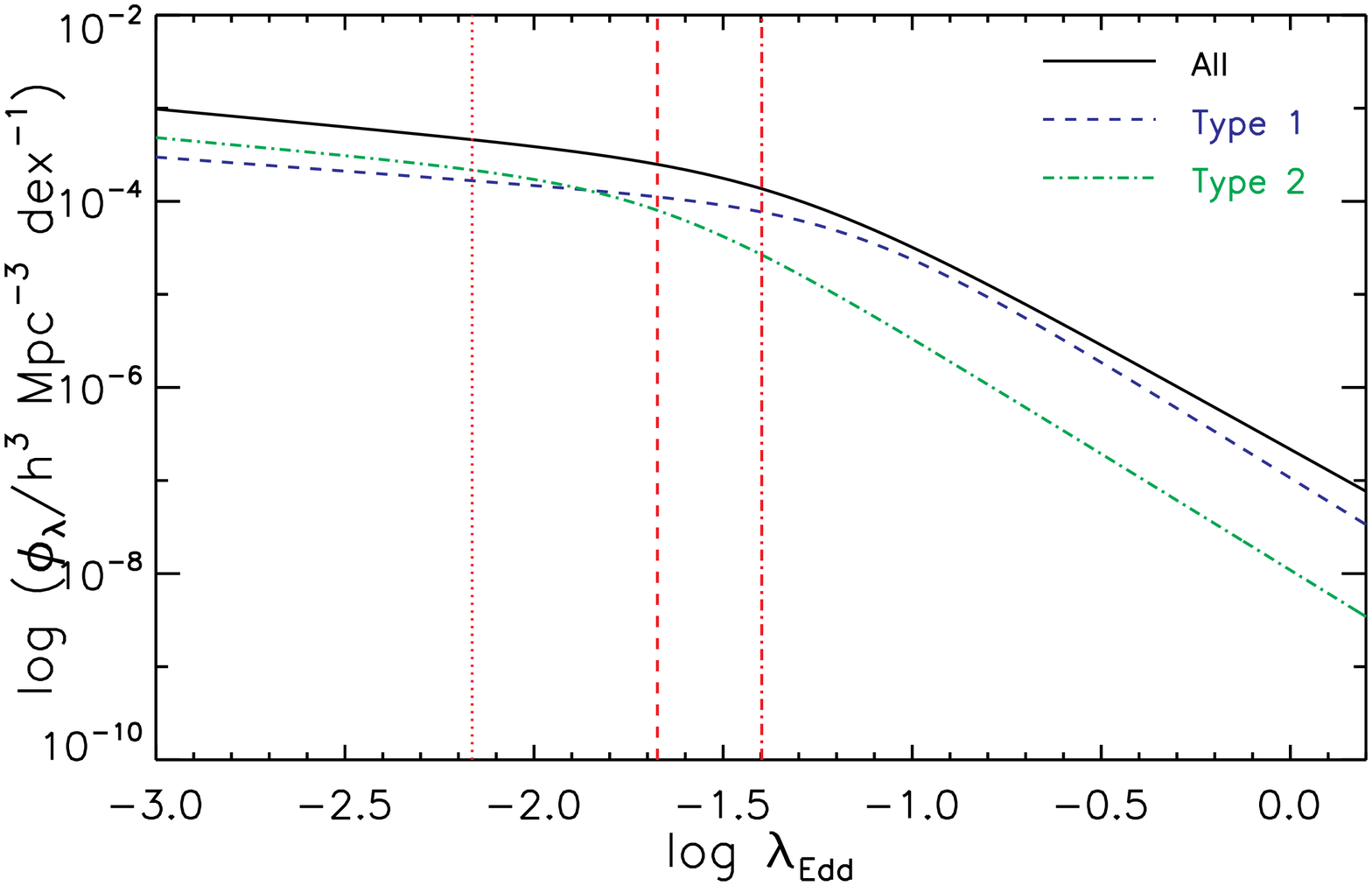}
 \par \smallskip
 \includegraphics[width=0.48\textwidth]{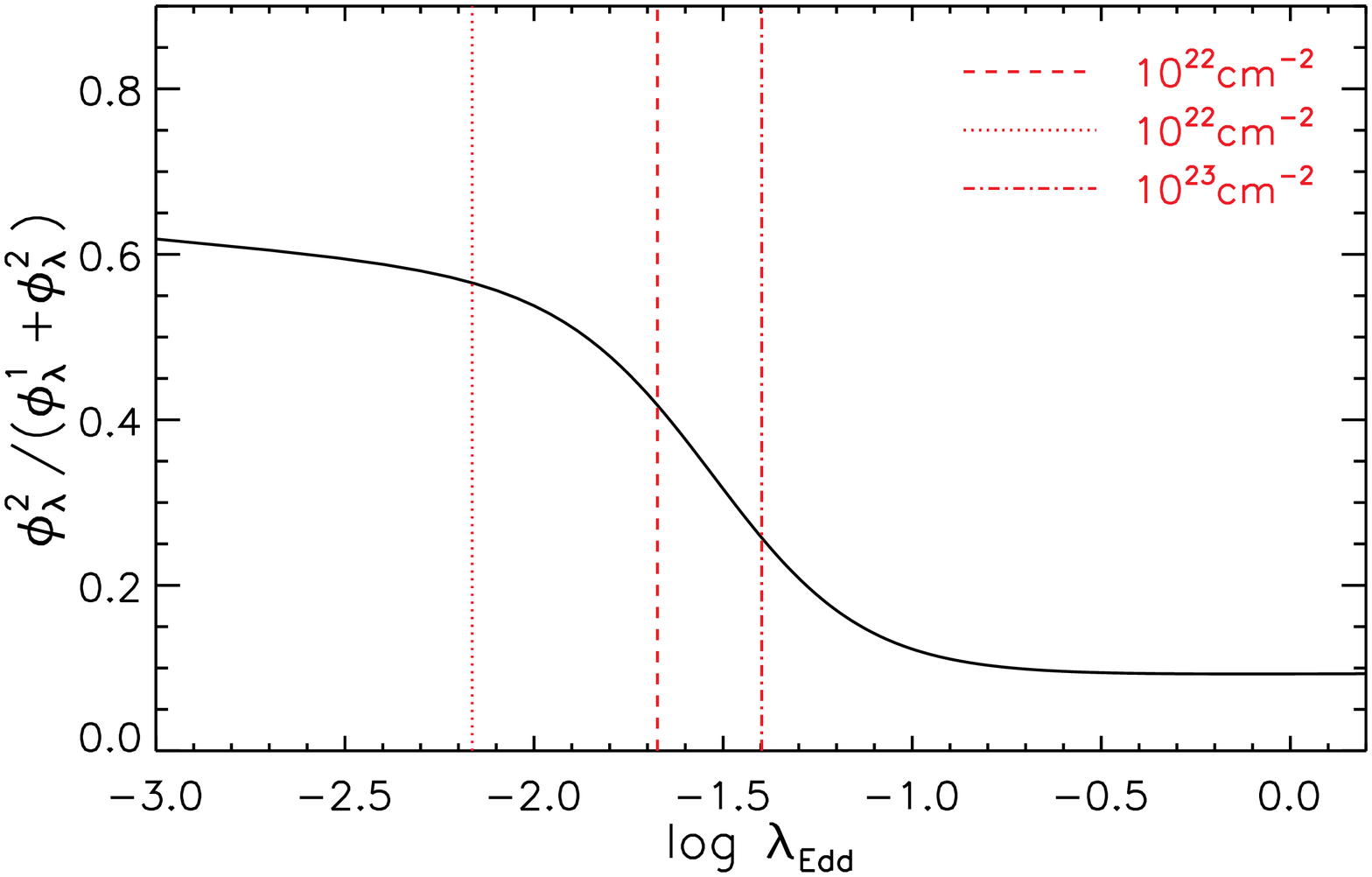}
  \caption{{\it Top panel:} Eddington ratio distribution function of type\,1 and type\,2 BASS AGN from \citet{Ananna:2022fg}. The vertical red lines show the expected Eddington limit for dusty gas with $\log (N_{\rm H}/\rm cm^{-2})\simeq 22$ (dashed line from \citealp{Fabian:2006lq,Fabian:2008hc,Fabian:2009ez}; dotted line from \citealp{Ishibashi:2018ti}) and $\log (N_{\rm H}/\rm cm^{-2})\simeq 23$ (dot-dashed line from \citealp{Ishibashi:2018ti}, see also \citealp{Venanzi:2020gx}). {\it Bottom panel}: ratio between the Eddington-ratio distribution function of type\,2 AGN and that of the whole AGN population, which is a proxy of the fraction of obscured sources and of the covering factor of the obscuring material.}
\label{fig:ERDF}
\end{figure}

\section{Radiation-regulated growth of supermassive black holes}\label{sec:RRGWTH}

\subsection{Luminosity and Eddington ratio distribution functions of nearby AGN}

Luminosity and Eddington ratio distribution functions can provide important insights on the lifetime of the different phases of SMBH growth. While AGN luminosity functions (LFs) have been studied in detail in different bands and at various redshifts (e.g., \citealp{Ueda:2003qf,Ueda:2014ix,Nagar:2005zr,Caputi:2007kl,Hopkins:2007mt,Paltani:2008bx,Aird:2010xo,Ross:2013va,Buchner:2015ve}), studies of the Eddington ratio distribution function (ERDF, e.g., \citealp{Kollmeier:2006eu,Greene:2007it,Aird:2012lq,Kelly:2013rl,Schulze:2015ru,Caplar:2015kq,Bongiorno:2016bh,Weigel:2017aa,Ananna:2022fg}) are still relatively scarce. Early efforts to study the ERDF were based on high-luminosity AGN at $z<0.3$ from the Hamburg/ESO Survey \citep{Schulze:2010uc}, and on $1 < z < 2$ AGN from different optical surveys \citep{Schulze:2015ru}. However, these samples were focussed on unobscured, high-luminosity AGN and did not provide a full picture of the SMBH growth. Some other studies have used the stellar mass as an indicator of the black hole mass, and derived ERDFs including for lower-luminosity, obscured systems (e.g., \citealp{Georgakakis:2017fs,Aird:2018ei}). Recently, studying nearby hard X-ray selected AGN from BASS, \cite{Ananna:2022fg} calculated the ERDF of AGN in the local Universe in the $10^{-3}<\lambda_{\rm Edd}<1$ range for both type\,1 and type\,2 AGN (top panel of Fig.\,\ref{fig:ERDF}), finding that the shape of the ERDF is independent of the AGN black hole mass. \cite{Ananna:2022fg} found that the total ERDF can be well reproduced by a double power-law with a break at $\log \lambda_{\rm Edd}^{*}=-1.34\pm0.07$. A similar value was inferred by \cite{Weigel:2017aa}, who proposed that the growth of radiatively efficient (i.e., X-ray detected) and inefficient (i.e., radio detected) AGN could each have universal ERDFs, which can reproduce both the black hole mass function and the AGN luminosity function. \cite{Weigel:2017aa} found that the ERDF of local, radiatively efficient AGN has a break between $\log \lambda_{\rm Edd}^{*}=-1.57$ and $\log \lambda_{\rm Edd}^{*}=-1.11$. 

As expected, the ratio between the ERDF of type\,2 and that of the whole AGN population (bottom panel of Fig.\,\ref{fig:ERDF}) shows a similar trend to that observed for the $f_{\rm obs}$--$\lambda_{\rm Edd}$ relation \citep{Ricci:2017rn}. The break in the ERDF of local type\,2 AGN ($\log \lambda_{\rm Edd}^{*}=-1.66^{+0.09}_{-0.06}$) found by \citet{Ananna:2022fg} is consistent with the Eddington limit for dusty gas, while that of type\,1 AGN is found at higher Eddington ratios ($\log \lambda_{\rm Edd}^{*}=-1.15^{+0.09}_{-0.05}$). Considering the strong anisotropy of the radiation produced in the accretion disk (e.g., \citealp{Kawaguchi:2010qc}), it is possible that this value is associated with the action of infrared radiation trapping in the optically thick material located along the plane of the accretion disk, which was not blown away at $\log \lambda_{\rm Edd}\sim -2$.

\begin{figure}
 \centering
 \includegraphics[width=0.48\textwidth]{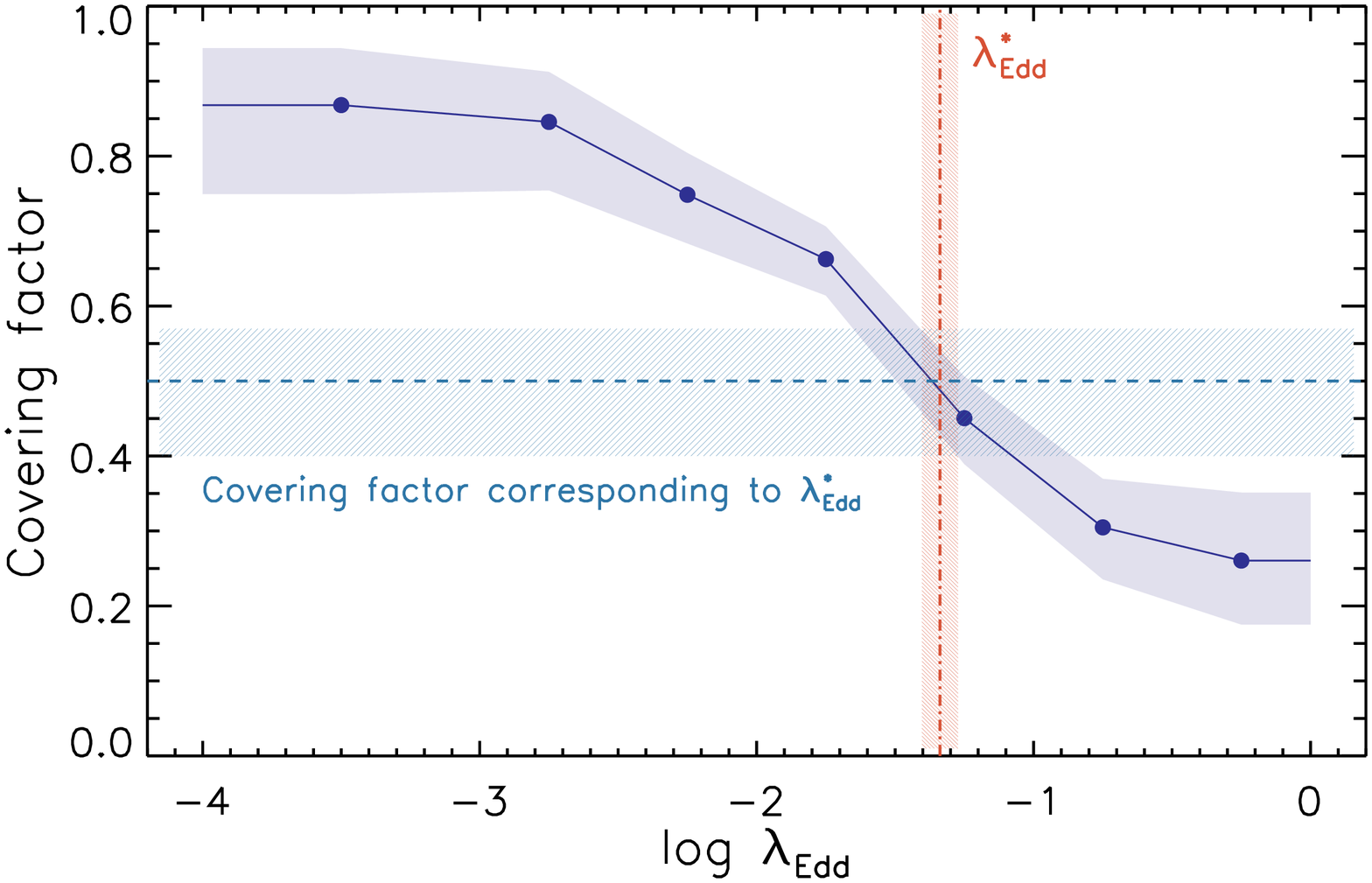}
 \par\medskip
 \includegraphics[width=0.48\textwidth]{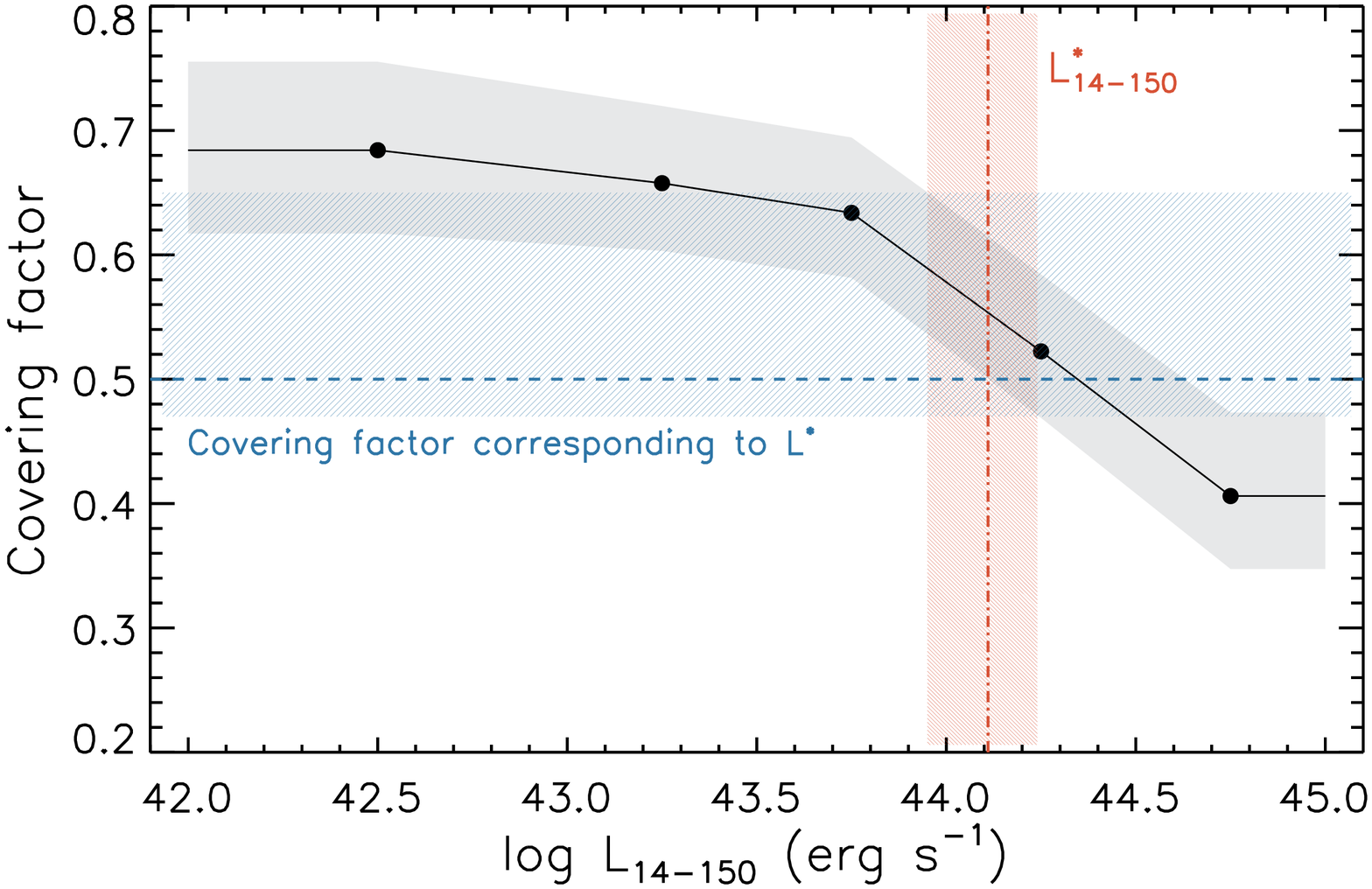}
  \caption{{\it Top panel:} Covering factor of obscuring material versus Eddington ratio for the objects in our hard X-ray selected sample with $-4 \leq \log \lambda_{\rm Edd}< 0$. The covering factor was obtained by considering the fraction of obscured sources, as in Fig.\,\ref{fig:FobsVsEdd}, including the Compton-thick fraction from \citet{Ricci:2017rn}, and normalising to unity in the $20 \leq \log (N_{\rm H}/\rm cm^{-2})< 25$ interval. The red vertical line shows the break in the Eddington ratio distribution function ($\lambda_{\rm Edd}^*$), while the hatched vertical area shows the uncertainty on $\lambda_{\rm Edd}^*$ \citep{Ananna:2022fg}. The blue hatched horizontal area represents the observed interval of the covering factor overlapping with $\lambda_{\rm Edd}^*$. The blue dashed line corresponds to $f_{\rm obs}=50\%$. The value of $\lambda_{\rm Edd}^*$ corresponds to the Eddington ratio where the transition between most of the sources being obscured ($f_{\rm obs}>50\%$) to most of the sources being unobscured ($f_{\rm obs}<50\%$) is observed. {\it Bottom panel:} same as the top panel for the 14--150\,keV luminosity. The red vertical line shows the break in the luminosity function ($L_{\rm 14-150}^*$), while the hatched vertical area shows its uncertainty \citep{Ananna:2022fg}.   }
\label{fig:FobsVsEddratio_breakERDF}
\end{figure}

\subsection{A radiation-regulated model for the growth of nearby SMBHs}

The value of $\lambda_{\rm Edd}^{*}$ inferred by \citet{Ananna:2022fg} for the whole AGN population lies in the range at which we find that BASS AGN transitions from being mostly obscured ($f_{\rm obs}>50\%$) to mostly unobscured ($f_{\rm obs}<50\%$), as illustrated in the top panel of Figure\,\ref{fig:FobsVsEddratio_breakERDF}. A similar trend is observed when considering the X-ray luminosity (bottom panel of Figure\,\ref{fig:FobsVsEddratio_breakERDF}), with the break ($L_{14-150}^*$) in the LF\footnote{\citet{Ananna:2022fg} report the luminosity in the 14--195\,keV band, which was converted into the 14--150\,keV luminosity adopted here assuming a power-law continuum with a photon index of $\Gamma=1.8$, consistent with the median of {\it Swift}/BAT AGN \citep{Ricci:2017pm}.} being consistent with the transition between an AGN being mostly obscured to being mostly unobscured. Most of the SMBH growth in the local Universe should occur around the break in the LF (e.g., \citealp{Hopkins:2005yw,Hopkins:2009yb}), which corresponds, for our sample of nearby AGN, to the phase during which the accreting SMBH is surrounded by large quantities of gas and dust.  Since the number density of AGN strongly decreases at $\lambda_{\rm Edd}\gtrsim \lambda_{\rm Edd}^{*}$, this suggests that, in removing the obscuring material, radiation pressure from the AGN also depletes the reservoir of the fuelling material, thus regulating accretion onto the SMBH, and causing the shorter lifetime of AGN accreting at high Eddington ratios.

\begin{figure}
\centering
 %% 1st image
 %% 2nd image
\includegraphics[width=0.48\textwidth]{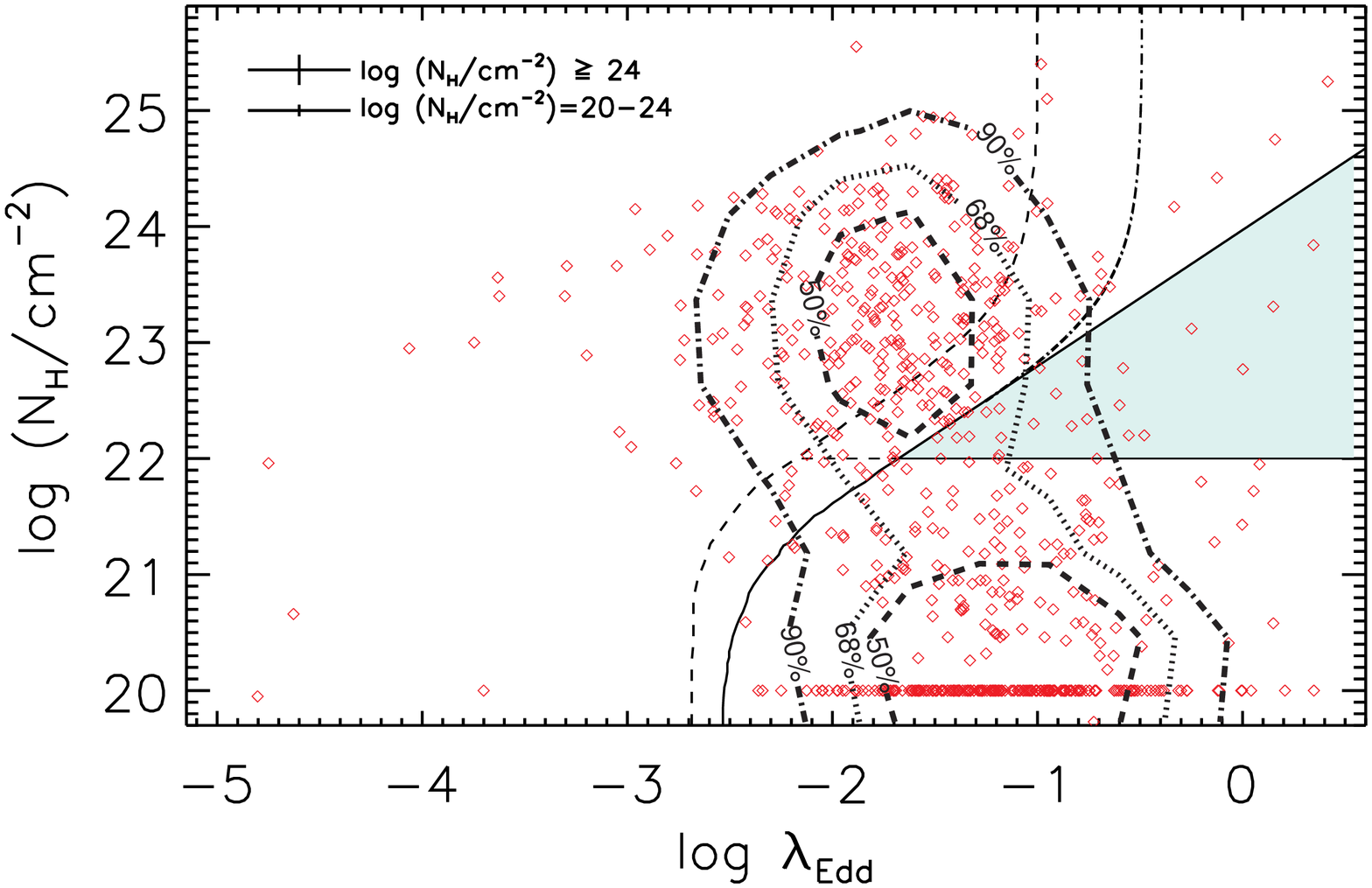}
\par \medskip
\includegraphics[width=0.48\textwidth]{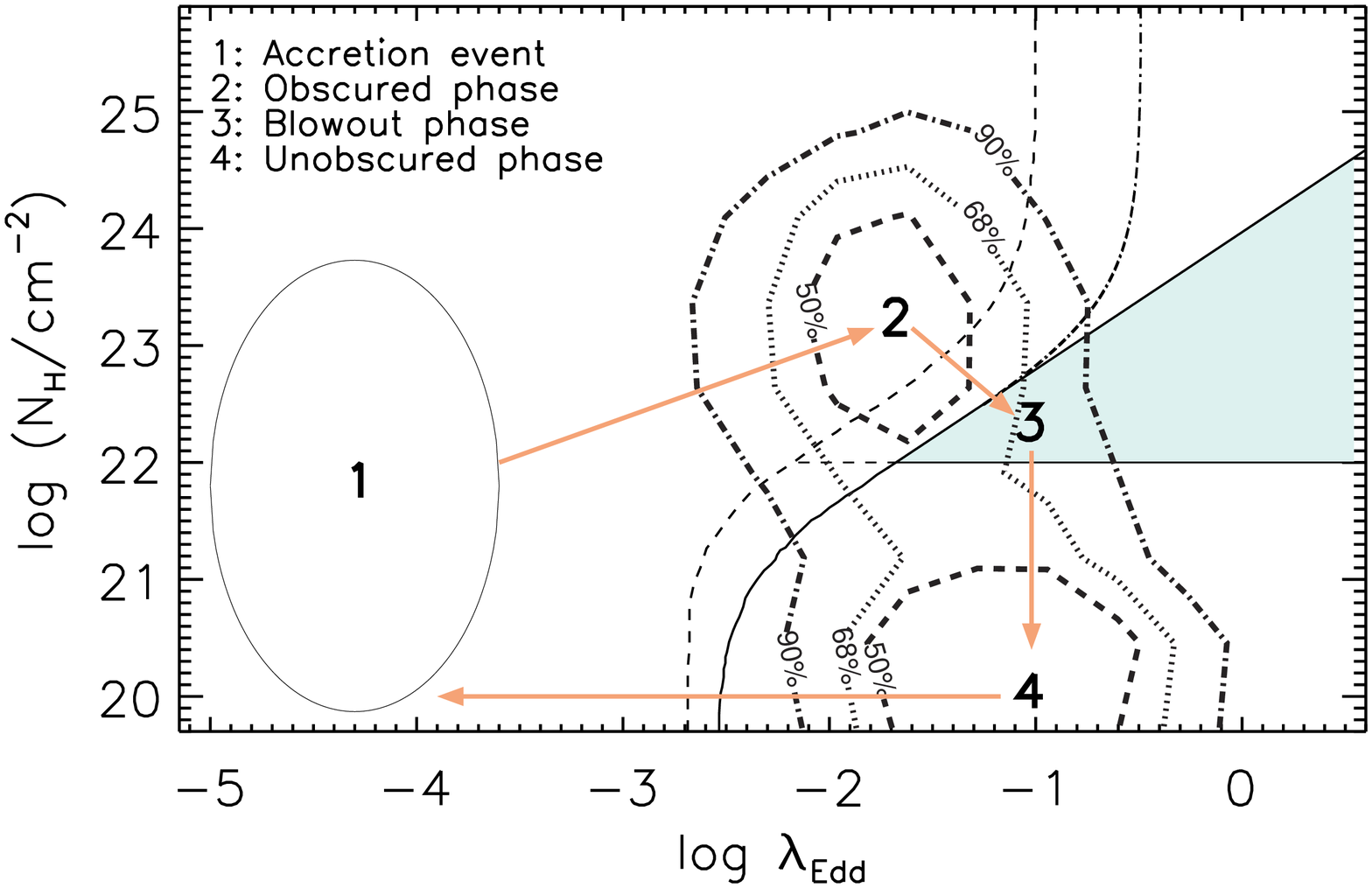}
% %% caption
% \begin{minipage}[t]{1\textwidth}
  \caption{{\it Top panel:} BASS AGN (red diamonds) in the $N_{\rm H}$--$\lambda_{\rm Edd}$ plane. The dashed, dotted and dot-dashed black lines represent the 50\%, 68\% and 90\% number density contours in the $10^{-3}\leq \lambda_{\rm Edd}\leq 1$ range. The black continuous line represents the effective Eddington limit for dusty gas reported in \citet{Fabian:2009ez}. The horizontal line at $\log (N_{\rm H}/\rm cm^{-2})=22$ represents the assumed maximum contribution to $N_{\rm H}$ of gas from the host galaxy. The colored surface represents the blowout region. The dash-dotted line shows the effective Eddington limit when including infrared radiation trapping \citep{Ishibashi:2018ti}, adapted to the values of  \citet{Fabian:2009ez}, similarly to what was done by \citet{Lansbury:2020mc}. The dashed line represents the effective Eddington limit for dusty gas reported by \citeauthor{Ishibashi:2018ti} (\citeyear{Ishibashi:2018ti}; see also \citealp{Venanzi:2020gx}). The typical uncertainties are reported in the top left corner for Compton-thin and CT AGN. Sources for which $N_{\rm H}$ was too low to be constrained were assigned $\log (N_{\rm H}/\rm cm^{-2})=20$. The low number of low $\lambda_{\rm Edd}$ points is due to relatively low sensitivity of {\it Swift}/BAT, which does not allow it to detect low-luminosity AGN. {\it Bottom panel:} a schematic of the radiation-regulated growth model outlined here: an accretion event (1) leads to an increase of the Eddington ratio and typical column density of an AGN (2), which would be preferentially observed as an obscured or type\,2 source, due to the large covering factor of the obscuring material. As the Eddington ratio increases above the effective Eddington limit for dusty gas, the AGN spends a short time in the blowout region (3), before its covering factor decreases, and it is mostly observed as an unobscured or type\,1 source (4). Once most of the material has been accreted, or blown away by infrared radiation trapping, the source moves back to having low values of $N_{\rm H}$ and $\lambda_{\rm Edd}$. The lines are the same as in the top panel.}
\label{fig:Evolution}
% \end{minipage}
\end{figure}

The differences in the Eddington ratio distributions of obscured and unobscured AGN, as well as the overall shape of the ERDF and LF, can be interpreted in the framework of an evolutionary model, in which radiation-pressure, besides shaping the close environment of SMBHs, also regulates their growth. A schematic of this radiation-regulated growth model of SMBHs, which is an extension of the radiation-regulated unification model \citep{Ricci:2017rn}, is shown in Figure\,\ref{fig:Evolution}: an accretion event (1) increases the Eddington ratio, typical column density and covering factor of the circumnuclear obscuring material of an AGN (2). Due to the large covering factor of the circumnuclear obscuring material, an AGN in this stage would be preferentially observed as obscured by an observer at a random inclination angle. As the Eddington ratio increases due to the large amount of fuel available, the source eventually reaches the effective Eddington limit for dusty gas. It then spends a short time in the blowout region (3), with the covering factor of the circumnuclear obscuring material decreasing rapidly, until most of the circumnuclear obscuring material is blown away, and the AGN is more likely observed as a relatively unobscured source (4). This would leave mostly optically-thick material located near the plane of the disk, in agreement with the finding that the covering factor of the CT gas does not change significantly with Eddington ratio \citep{Ricci:2017rn}. Once most of the material has been accreted and/or has been blown away by radiation pressure and IR trapping, the source would transition to low $N_{\rm H}$ and $\lambda_{\rm Edd}$. 

The timescales of this process will be discussed in detail, using the ERDF, in a forthcoming companion paper (Ananna et al. 2022b; see also \citealp{Lansbury:2020mc} and \citealp{Jun:2021rr} for recent discussion on the blowout timescales). It should be noted that in Fig.\,\ref{fig:Evolution} we assumed a maximum contribution of the host galaxy to the X-ray obscuration of $\log (N_{\rm H}/\rm cm^{-2})\sim 22$, i.e. $\sim 1$\,dex higher than the angle-averaged value obtained for the Milky way (e.g., \citealp{Willingale:2013lo}), although this value could be larger for high stellar masses (e.g., \citealp{Buchner:2017jx}) and could increase significantly at higher redshifts (e.g., \citealp{Banerji:2012du,Assef:2015zr,LaMassa:2016mn,Gilli:2022pu}; see \S\ref{sect:largescaleobs}). The accretion event could be either associated with secular processes (e.g., \citealp{Davies:2007qo}) or with mergers (e.g., \citealp{Blecha:2018gt}). Both mechanisms could be at play in our AGN population: {\it Swift}/BAT AGN tend to be hosted by gas-rich spiral galaxies (e.g., \citealp{Koss:2011rs}) and to have high gas fractions \cite{Koss:2021zw}, and some of them can be found in galaxy mergers (e.g., \citealp{Koss:2018cw}).

According to the model proposed here, the obscuring material of objects at the end of the cycle, with low Eddington ratios ($\log \lambda_{\rm Edd}\lesssim-4$), would be expected to have lower covering factor and column density. While the statistics available are still small, this could be reflected in the lower fraction of obscured sources observed in the $-4.8 \leq \log \lambda_{\rm Edd} \leq -4$ range (\citealp{Ricci:2017rn}, Fig.\,\ref{fig:FobsVsEdd}). These very low-Eddington ratio AGN typically show only faint X-ray reflection features (e.g., \citealp{Ptak:2004uf}), and recent modelling of their broad-band X-ray spectra has also shown that their integrated column densities appear to be significantly lower than those of more rapidly accreting SMBHs (e.g., \citealp{Ursini:2015qy,Diaz:2020zr}). This might not be the case for all low $\lambda_{\rm Edd}$ AGN, since some of them might be in the process of starting a new cycle (i.e. transitioning from step one to two in Fig.\,\ref{fig:Evolution}). Low-luminosity AGN (e.g., \citealp{Ho:2008ly,Ho:2009dl}) are still expected to produce feedback, preferentially through the kinetic mode (e.g., \citealp{Weinberger:2017te}), which would however mostly heat up the gas on Galactic scales, and not strongly affect the circumnuclear environment of the SMBH.

\subsection{The role of large-scale obscuration at high redshift}\label{sect:largescaleobs}
 
At higher redshifts and luminosities than those probed here, it seems to be rather common for sources accreting at very high $\lambda_{\rm Edd}$ to be obscured. This is shown, for example, by the relatively high number density of Hot Dust Obscured Galaxies (Hot DOGs) which are found to have large [$\log (N_{\rm H}/\rm cm^{-2})>23$] column densities (e.g., \citealp{Stern:2014kx,Ricci:2017ic,Zappacosta:2018wr,Vito:2018du}) compared to AGN with similar luminosities \citep{Assef:2015zr}. Red quasars, which host obscured AGN (e.g., \citealp{Banerji:2012du,LaMassa:2016mn,Glikman:2017vu}), have also been found to reside in the blowout region of the $N_{\rm H}$--$\lambda_{\rm Edd}$ diagram \citep{Glikman:2017ug,Lansbury:2020mc}. Moreover, many of these objects show evidence of powerful outflows in their optical/UV spectra (e.g., \citealp{Yi:2022pa}), suggesting they are in the process of driving material away from the nucleus \citep{Temple:2019pr, Lansbury:2020mc}. \citet{Jun:2021rr} recently studied a large sample of infrared and submillimeter-bright obscured quasars with bolometric luminosities $L_{\rm Bol}\gtrsim 10^{46}\rm\,erg\,s^{-1}$, which included red quasars, Hot DOGs, ultra-luminous infrared galaxies and submillimeter galaxies. \citet{Jun:2021rr} showed that most of these objects are found in the blowout region, and behave differently from lower luminosity, local AGN. In some of these objects, the AGN might be in the brief phase during which it is expelling dusty circumnuclear gas, and cleaning up its environment. This behavior in IR-bright and submm galaxies could also be related to the high fraction of mergers found in these objects (e.g., \citealp{Urrutia:2008vn,Glikman:2015lk,Fan:2016vy,Diaz-Santos:2018gj}). Theoretical studies have shown that the merger process can be very efficient in moving gas into the inner hundreds of parsecs of galaxies, thus feeding and obscuring the SMBH (e.g., \citealp{Hopkins:2008xr,Blecha:2018gt,Kawaguchi:2020qb}). In agreement with this, it has been found that the typical column density and covering factor of the obscuring material is very high in galaxies undergoing mergers, particularly during the final stages (\citealp{Ricci:2017aa,Ricci:2021oz}; see also \citealp{Satyapal:2014oq,Kocevski:2015zr,Yamada:2021vz}). Part of this material could be located outside the sphere of influence of the SMBH, so that the main parameter determining the effect of radiation pressure would be the luminosity, rather than the Eddington ratio. Moreover, it has been argued that the presence of stars in the dusty gas clouds would hinder the effect of radiation pressure \citep{Fabian:2009ez}, since they would provide an additional gravitational pull, allowing the obscuring material to survive to higher $\lambda_{\rm Edd}$. Therefore, in the case of major merger-induced accretion the radiation-regulated growth model illustrated in Figure\,\ref{fig:Evolution} could reach higher column densities (e.g., \citealp{Ricci:2017aa,Ricci:2021oz}) and accretion rates (e.g., \citealp{Treister:2012vn}). 

The ISM could also play a role in obscuring the AGN at redshifts higher than those probed by our study. Using ALMA observations, \cite{Gilli:2022pu} recently showed that the column density of the ISM towards the nucleus of $z > 3$ galaxies is typically $> 100$ times larger than at $z\sim 0$, and it may reach CT values at $z \gtrsim 6$. This, combined with the larger number of AGN in mergers at higher redshifts with respect to $z\simeq 0$ (e.g., \citealp{Mortlock:2013is,Whitney:2021eu}), could explain the increase of the fraction of obscured sources with redshift (e.g., \citealp{La-Franca:2005ec,Treister:2006se,Ueda:2014ix,Aird:2015gf,Buchner:2015ve,Vito:2018ot,Avirett-Mackenzie:2019ut}), and in particular the high fraction of AGN with $\log (N_{\rm H}/\rm cm^{-2})\gtrsim 23$ at $z>3$ \citep{Vito:2018ot}. Since the ISM material is outside the sphere of influence of the SMBH, in these high-z objects the relation between radiative AGN feedback and the obscuring material would be regulated by luminosity, and not by Eddington ratio. This would allow AGN to reach higher luminosities before they are able to remove the obscuring gas. If part of the obscuring material is associated with the AGN feeding one would expect the break in the ERDF to shift to higher values for higher redshifts (e.g., \citealp{Schulze:2015ru,Caplar:2015kq}), similarly to what is observed for the break in the luminosity function (e.g., \citealp{Ueda:2014ix}).

\section{Summary and conclusions}

We studied here the relation between AGN obscuration and the Eddington ratio using a highly complete sample of 681 nearby X-ray selected AGN with black hole mass measurements from BASS, with the goal of improving our understanding of the relation between obscuration, radiation pressure and SMBH growth. Our main findings are:

\begin{itemize}
\item Thanks to the significantly larger sample, we confirm with a higher statistical significance the results obtained by \citet{Ricci:2017rn}, namely that the fraction of obscured sources decreases sharply at $\lambda_{\rm Edd}\gtrsim 10^{-2}$ (Fig.\,\ref{fig:FobsVsEdd}), corresponding to the Eddington limit for dusty gas \citep{Fabian:2006lq,Fabian:2008hc,Fabian:2009ez}. This suggests that radiative feedback can efficiently remove the obscuring material around SMBHs.
\item Using the large BASS dataset, we find a strong increase in the fraction of sources with $\log (N_{\rm H}/\rm cm^{-2})<21$ at  $\lambda_{\rm Edd}\gtrsim 10^{-2}$, consistent with the idea that obscured sources become fully unobscured rapidly once they accrete above the effective Eddington limit for dusty gas (top left panel of Fig.\,\ref{fig:FobsVsEdd_NHbins}). 

\item The fraction of sources with $21 \leq \log (N_{\rm H}/\rm cm^{-2})<22$ does not change with $\lambda_{\rm Edd}$, and is stable at $\sim 15-20\%$, suggesting that most of the (neutral or weakly-ionized) obscuration in these AGN is produced by gas in their host galaxies (top right panel of Fig.\,\ref{fig:FobsVsEdd_NHbins}).
\item The fraction of sources with $22 \leq \log (N_{\rm H}/\rm cm^{-2})<23$ decreases rapidly at $\lambda_{\rm Edd}\gtrsim 10^{-2}$ (bottom left panel of Fig.\,\ref{fig:FobsVsEdd_NHbins}), consistent with the radiation-regulated unification model \citep{Ricci:2017rn}. Interestingly, the fraction of sources with $23 \leq \log (N_{\rm H}/\rm cm^{-2})<24$ also decreases at a similar Eddington ratio (bottom right panel of Fig.\,\ref{fig:FobsVsEdd_NHbins}). This could be due to the effect of infrared radiation trapping (e.g., \citealp{Ishibashi:2018ti,Venanzi:2020gx}), which is expected to lead material with $\log (N_{\rm H}/\rm cm^{-2})\gtrsim 23$ to evaporate at $\lambda_{\rm Edd}$ below the expected Eddington limit for dusty gas with the same column density.
\item Using the ratio between the ERDF of type\,2s and the ERDF of type\,1 and type\,2 AGN \citep{Ananna:2022fg} we recover a similar relation between the covering factor and the Eddington ratio as that found using the fraction of obscured sources (Fig.\,\ref{fig:ERDF}; see also \citealp{Ricci:2017rn}). 
\item The breaks in the ERDF ($\log \lambda_{\rm Edd}^{*}=-1.34\pm0.07$) and LF ($\log L_{14-150}^*=44.11^{+0.13}_{-0.16}$) of all BAT AGN \citep{Ananna:2022fg} are found where AGN transition from having most of their sky covered by obscuring material to having most of its sky devoid of absorbing material (Fig.\,\ref{fig:FobsVsEddratio_breakERDF}). This implies that most of the SMBH growth in the local Universe would happen when the AGN is covered by a large fraction of gas and dust. The fact that AGN with Eddington ratios above $\lambda_{\rm Edd}^{*}$, which have low covering factors of the obscuring material, are rarer could be associated with the lower amount of accreting material available in their surroundings, which would lead them to spend a relatively short time in this phase.
\item We suggest that these results could be explained with a radiation-regulated growth model for AGN (see Fig.\,\ref{fig:Evolution}), in which (nearby) accreting SMBHs move in the $N_{\rm H}$--$\lambda_{\rm Edd}$ plane during their life cycle. The growth episode starts with the AGN mostly unobscured and accreting at low $\lambda_{\rm Edd}$. As the SMBH receives its fuel, its $\lambda_{\rm Edd}$, $N_{\rm H}$ and covering factor increase. At this stage, an observer, with a randomly selected inclination angle with respect to the obscuring material, would preferentially observe the source as an obscured/type\,2 AGN. When $\lambda_{\rm Edd}$ increases above the Eddington limit for dusty gas, the AGN starts to rapidly expel the obscuring material, which leads to a rapid decrease of its covering factor and typical $N_{\rm H}$. The AGN is now observed typically as unobscured/type\,1, with only CT material along the equatorial plane left. As the material is depleted, either by accretion or by the effect of infrared radiation trapping due to the increasing $\lambda_{\rm Edd}$, the SMBH goes back to a quiescent phase.

\end{itemize}

In a parallel and complementary BASS analysis (Ananna et al. 2022b) we will use the ERDF of nearby AGN to constrain the timescales of the different stages of SMBH growth. At redshifts higher than those probed here, large-scale material associated to the host galaxy might play an important role in obscuring AGN (e.g., \citealp{Gilli:2022pu}), and the main parameter determining the effect of radiation pressure could be the luminosity, rather than the Eddington ratio. Future studies of the evolution with redshift of the $f_{\rm obs}$--$\lambda_{\rm Edd}$ relation, and of the $N_{\rm H}$--$\lambda_{\rm Edd}$ diagram, will help shedding light on how the relation between AGN obscuration and radiative feedback changes over cosmic time.

\facility{Swift}

\appendix

\section{The effect of black hole mass and luminosity on the obscuration--Eddington ratio relation}\label{sec:appendix1}

\begin{figure*}
\centering
 %% 1st image
 %% 2nd image
\includegraphics[width=0.95\textwidth]{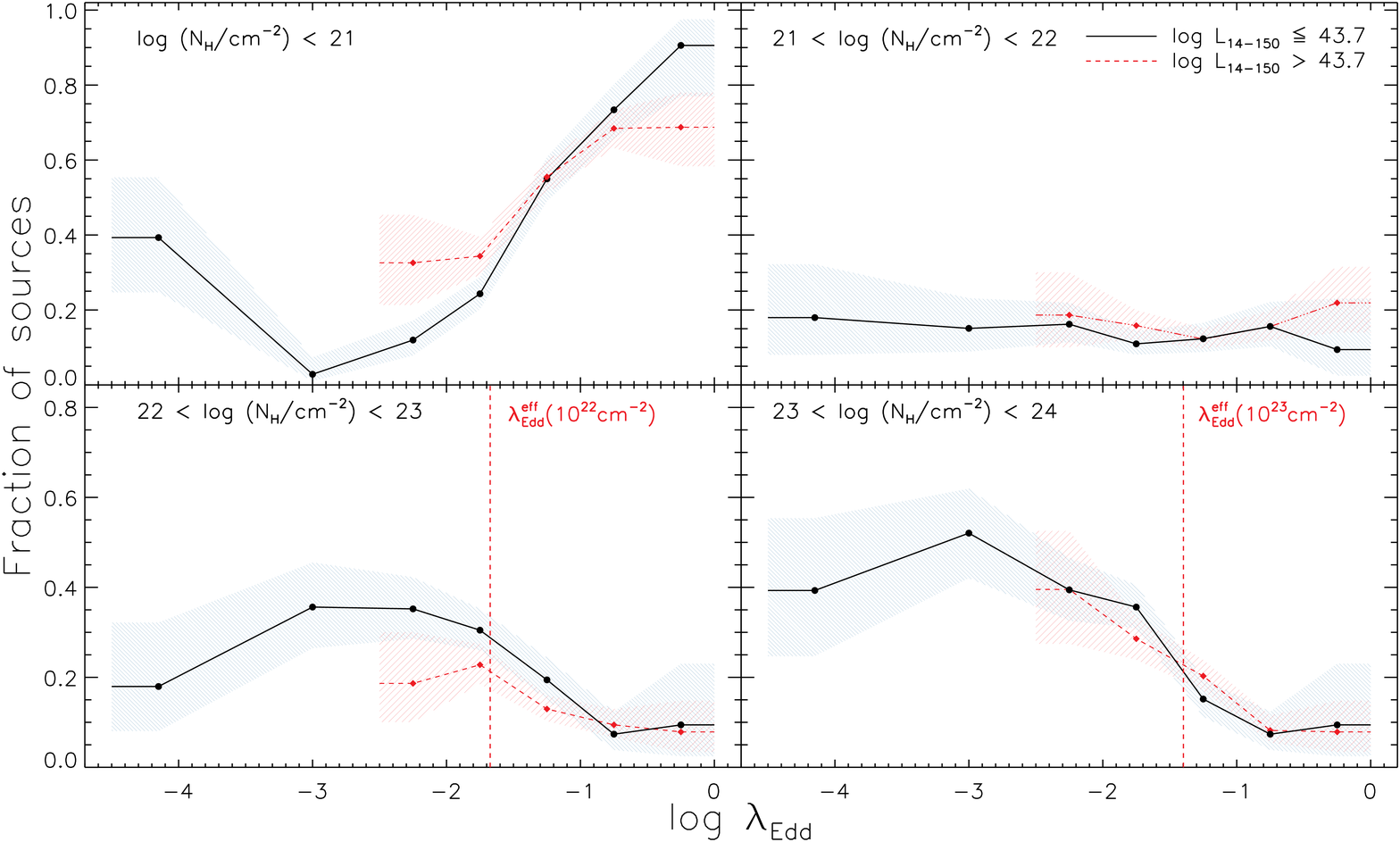}
\par \medskip
\includegraphics[width=0.95\textwidth]{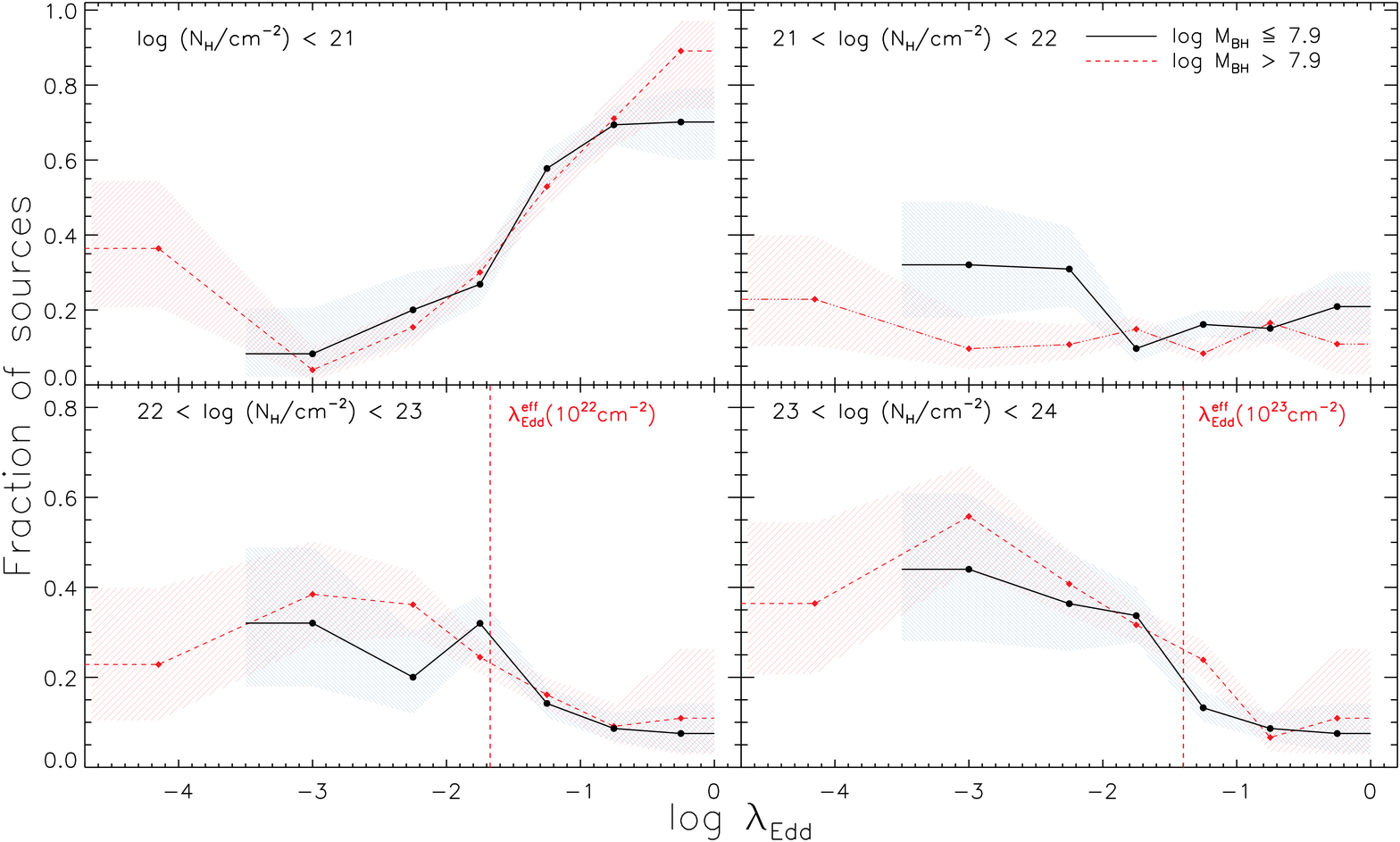}
% %% caption
% \begin{minipage}[t]{1\textwidth}
  \caption{{\it Top panel:} fraction of sources with column densities in a given range versus the Eddington ratio for AGN in two different bins of intrinsic 14--150\,keV luminosity (in units of $\rm erg\,s^{-1}$. The red dashed lines show the expected Eddington limit for dusty gas with $\log (N_{\rm H}/\rm cm^{-2})\simeq 22$ (bottom left panel; \citealp{Fabian:2006lq,Fabian:2008hc,Fabian:2009ez}) and $\log (N_{\rm H}/\rm cm^{-2})\simeq 23$ (bottom right panel; \citealp{Ishibashi:2018ti,Venanzi:2020gx}). The latter value of the effective Eddington limit includes the contribution from IR radiation trapping. {\it Bottom panel:} same as top panel, but with AGN divided into two different ranges of black hole mass (in units of Solar masses).}
\label{fig:FobsVsEdd_NHbins_lumMBH}
% \end{minipage}
\end{figure*}

In the top and bottom panels of Fig.\,\ref{fig:FobsVsEdd_NHbins_lumMBH} we show the relation between the fraction of sources with given column density and the Eddington ratio. The sources are divided into two different ranges of 14--150\,keV intrinsic luminosity (top panels) and black hole mass (bottom panels), showing that these two parameters do not play a significant role.

\acknowledgments
We thank the referee for their positive and helpful report, which helped us to improve the quality of the paper. We thank G. Lansbury for sharing some of the data of his 2020 paper with us. We acknowledge support from the National Science Foundation of China (11721303, 11991052, 12011540375) and the China Manned Space Project (CMS-CSST-2021-A04, CMS-CSST-2021-A06) (LH), Fondecyt Iniciacion grant 11190831 (CR) and ANID BASAL project FB210003 (CR, FEB, ET); NASA through ADAP award NNH16CT03C (MK); the Israel Science Foundation through grant number 1849/19 (BT); the European Research Council (ERC) under the European Union's Horizon 2020 research and innovation program, through grant agreement number 950533 (BT); Fondecyt fellowship No. 3220516 (MT); the Korea Astronomy and Space Science Institute under the R\&D program(Project No. 2022-1-868-04) supervised by the Ministry of Science and ICT (KO);  the National Research Foundation of Korea (NRF-2020R1C1C1005462) (KO); RIN MIUR 2017 project ''Black Hole winds and the Baryon Life Cycle of Galaxies: the stone-guest at the galaxy evolution supper", contract 2017PH3WAT (FR); the ANID - Millennium Science Initiative Program - ICN12\_009 (FEB), CATA-Basal - ACE210002 (FEB, ET), FONDECYT Regular - 1190818 (FEB, ET) and 1200495 (FEB, ET), N\'ucleo Milenio NCN\_058 (ET), the Science Fund of the Republic of Serbia, PROMIS 6060916, BOWIE and by the Ministry of Education, Science and Technological Development of the Republic of Serbia through the contract No.~451-03-9/2022-14/200002 (MS).

\clearpage

\bibliography{radregulated_growth}
\bibliographystyle{aasjournal}

\end{document}